\documentclass[twocolumn,preprintnumbers,amsmath,amssymb,pra]{revtex4}

\usepackage{graphicx}
\usepackage{bm}

\begin{document}

\title{Cross-Kerr effective Hamiltonian for a non-resonant four-level atom}

\author{Gary F. Sinclair and Natalia Korolkova}
 \affiliation{School of Physics and Astronomy, University of St Andrews, North Haugh, St Andrews, KY16 9SS, Scotland}

\begin{abstract}
We derive a cross-Kerr type effective Hamiltonian for the four-level atom interacting with three electromagnetic fields in the N-configuration.  When the atom has relaxed into the ground state a cross-Kerr nonlinearity arises between two weak probe fields.  As a development on earlier work \cite{CKIFLAS} we show in general that the atom will also display a linear and self-Kerr response.  However, if certain resonance conditions are satisfied then the linear and self-Kerr interactions will vanish.  The electrical susceptibilities of the probe transitions are also explored and it is shown that a large, pure cross-Kerr nonlinearity can be generated with vanishing absorption of both probe fields.
\end{abstract}

\maketitle

\section{Introduction}
The exploration of nonlinear optics developed rapidly after the demonstration of the first laser by Maiman \cite{SORR} in 1960. However, the cross-Kerr nonlinearity was discovered much earlier in 1875 by the Scottish physicist, the Rev. John Kerr.  Originally, the refractive index experienced by a light ray propagating in the Kerr-medium was controlled by applying a strong DC electric field.  The field induces a birefringence in the material so that light polarised parallel to the electric field experiences an enhanced refractive index.  The Kerr-cell found several applications in early television recievers and high-speed photography.  

The advent of high-intensity laser light enabled the substitution of the DC field with an optical electromagnetic field.  The cross-modulation of the refractive index experienced by one electromagnetic field by the intensity of the other is called the optical (or AC) cross-Kerr effect.  This interaction has recently found many new applications: for instance, entanglement concentration \cite{ECCVQS, PECCVSL}, quantum teleportation \cite{CQTKN}, quantum state conversion \cite{QSCCKI} and quantum non-demolition measurements \cite{HEQNDD}.  It has also been proposed as a deterministic quantum logic gate \cite{WNOQC, CTXPM}.

In bulk materials, such as optical fibers, the cross-Kerr interaction is always accompanied by competing nonlinear effects, such as self-phase modulation and sum- and difference-frequency generation \cite{NLO}.   However, by utilising coherent atomic interactions, systems that generate a pure cross-Kerr interaction have been proposed \cite{GKNEIT, LXPMSCP, CKIFLAS} and experimentally demonstrated \cite{OLKN}.

\begin{figure}[ht]
  \center
  \includegraphics[width=8cm]{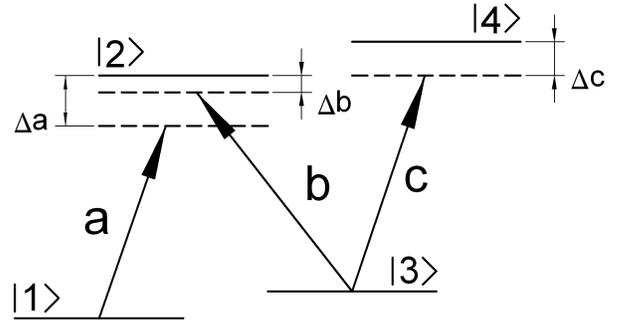}
  \caption{The four-level atomic system interacting with three electromagnetic field modes in the undressed atomic basis. $\Delta_x$ is the single photon detuning corresponding to the electromagnectic field mode ``x''.}
  \label{int:fls}
\end{figure}

Consider a system consisting of two electromagnetic field modes: mode `a' and mode `c'.  The Hamiltonian for the cross-Kerr interaction is given by
\begin{equation}
  \label{int:ckh}
  \hat{H}=\hbar K \hat{n}_a \hat{n}_c,
\end{equation}
where $K$ is the coupling strength and $\hat{n}_a$ and $\hat{n}_c$ are the photon-number operators.  The evolution determined by this Hamiltonian produces a cross-phase modulation.  Namely, when acting on a tensor product between two photon-number states ($\vert \psi(0) \rangle=\vert n_a \rangle \otimes \vert n_c \rangle$) we find that
\begin{equation}
  \vert \psi(t) \rangle = \exp(- i K n_a n_c t) \vert \psi(0) \rangle. \label{int:cke}
\end{equation}
In this paper, we show that a cross-Kerr type effective Hamiltonian (\ref{int:ckh}) can be generated in the four-level atom in the N-configuration (Fig. \ref{int:fls}).  The limit is considered where there are two weak probe fields (`a' and `c') and one strong pump (`b').  As a development from our previous publication \cite{CKIFLAS}, we remove the two-photon resonance condition between the modes `a' and `b'.  This greatly improves the model by showing how the strength of the cross-Kerr nonlinearity depends on the detunings of all three fields.  Moreover, we demonstrate the existance of imperfections in the cross-Kerr interaction, such as a linear and self-Kerr response.  The model also enables us to calculate the electrical susceptibilities from which the probe field absorptions can be calculated.  We remove the two-photon resonance condition by using non-degenerate time-independent perturbation theory in two variables to calculate the approximate eigenstates of the system.  The evolution of the radiatively stable ``ground'' eigenstate is calculated, from which an effective Hamiltonian for the relaxed atomic system is calculated.

\section{Model of the Atomic System}
We model a single atom interacting with three continuous-wave monochromatic electromagnetic fields (see Fig. (\ref{int:fls})).  Both the atom and the three fields are treated quantum mechanically.  All possible states of the atom are spanned by the four eigenstates $\vert 1 \rangle_A, \vert 2 \rangle_A, \vert 3 \rangle_A,$ and $\vert 4 \rangle_A$.  Similarly, the state of each electromagnetic field mode `x' can be expanded in the Fock basis $\{ \vert n_x \rangle$ : $n_x \in [0, \infty) \}$.

Therefore, the composite state of the atom and three fields system can be expanded in a basis consisting of the tensor product between the basis vectors of the individual components. 
\begin{equation}
  \label{moas:exp}
  \vert \psi \rangle = \sum^4_{i=1} \mathop{\mathop{\sum^{\infty}_{n_a=0}}_{n_b=0}}_{n_c = 0} c_{i, n_a, n_b, n_c} \vert i \rangle_A \otimes \vert n_a \rangle \otimes \vert n_b \rangle \otimes \vert n_c \rangle.
\end{equation}
Henceforth, we will omit the tensor product symbols.
\begin{equation}
 \vert i \rangle_A \otimes \vert n_a \rangle \otimes \vert n_b \rangle \otimes \vert n_c \rangle = \vert i, n_a, n_b, n_c \rangle.
\end{equation}
We assume that the electromagnetic fields couple to the atom by the electric-dipole interaction.  Considering only energy-conserving terms, the total Hamiltonian is written in the Schr\"{o}dinger picture as \cite{IQO}
\begin{equation}
  \hat{H}_{sp} = \sum^{4}_{i=1} E_i \hat{\sigma}_{i,i} + \mathop{\hbar \sum}_{k=\{a,b,c\}} \nu_k \hat{a}^{\dag}_k \hat{a}_k + g_k \hat{\sigma}_{k} \hat{a}_k + g^{*}_k \hat{\sigma}^{\dag}_k \hat{a}^{\dag}_k.
\end{equation}
Here,  $\nu_k$ is the angular frequency of the electromagnetic field mode `k'; $\hat{a}^{\dag}_k$ and $\hat{a}_k$ are the creation and annihilation operators and $g_k$ and $\hat{\sigma}_k$ are the coupling strengths and atomic-transition operators for the allowed electric-dipole transitions.
\begin{equation}
  g_a = g^{21}_a \qquad g_b = g^{23}_b \qquad g_c = g^{43}_c,
  \label{moas:g}
\end{equation}
\begin{equation}
  \hat{\sigma}_a = \vert 2 \rangle_{AA}\langle 1 \vert \qquad \hat{\sigma}_b = \vert 2 \rangle_{AA} \langle 3 \vert \qquad \hat{\sigma}_c = \vert 4 \rangle_{AA} \langle 3 \vert.
\end{equation}
It is convenient to transform the Hamiltonian into the interaction picture where the atomic levels are separated by the multi-photon detunings.  This is done by writing the non-coupling terms of the Hamiltonian in terms of the `conversion' operator invariants of the Hamiltonian.
\begin{equation}
  \hat{H}=\hbar \left (\delta_1 \hat{\sigma}_{22} + \delta_2 \hat{\sigma}_{33} + \delta_3 \hat{\sigma}_{44} \right ) + \mathop{\hbar \sum}_{k=\{a,b,c\}} g_k \hat{\sigma}_{k} \hat{a}_k + g^{*}_k \hat{\sigma}^{\dag}_k \hat{a}^{\dag}_k.
  \label{moas:hint}
\end{equation}
The multi-photon detunings are defined as
\begin{equation}
  \begin{array}{rcl}
    \delta_1 &=& \Delta_a, \\
    \delta_2 &=& \Delta_a - \Delta_b, \\
    \delta_3 &=& \Delta_a - \Delta_b + \Delta_c.
  \end{array}
  \label{moas:deltas}
\end{equation}
By considering the action of this Hamiltonian on the basis vectors from the set $\{ \vert i, n_a, n_b, n_c \rangle \}$, one can show that the system will evolve within a four-dimensional resonant-manifold \cite{ACPT}.  For instance, the state $\vert 1, n_a, n_b, n_c \rangle$ is only coupled to the states:
\begin{equation}
  \label{moas:rm}
  \begin{array}{c}
    \vert 2, n_a-1, n_b, n_c \rangle, \\
    \vert 3, n_a-1, n_b+1, n_c \rangle, \\
    \vert 4, n_a-1, n_b+1, n_c-1 \rangle.
  \end{array}
\end{equation}
Since every basis vector in the expansion (\ref{moas:exp}) can be written as a state belonging to a resonant manifold of the form (\ref{moas:rm}), we need only consider the behaviour of the system within one such prototype resonant manifold.

Henceforth we will assume that the state $\vert 1, n_a, n_b, n_c \rangle$ is denoted by $\vert 1 \rangle$ and
\begin{equation}
  \begin{array}{rcl}
    \vert 2 \rangle &=& \vert 2, n_a-1, n_b, n_c \rangle, \\
    \vert 3 \rangle &=& \vert 3, n_a-1, n_b+1, n_c \rangle, \\
    \vert 4 \rangle &=& \vert 4, n_a-1, n_b+1, n_c-1 \rangle.
  \end{array}
\end{equation}
When performing calculations it will be useful to express the Hamiltonian in matrix form.  We will make use of the canonical basis:
\begin{equation}
  \begin{array}{rclrcl}
    \vert 1 \rangle &=& \left [ \begin{array}{c}1\\0\\0\\0\end{array} \right ], &  \vert 2 \rangle &=& \left [ \begin{array}{c}0\\1\\0\\0\end{array} \right ], \\
    \vert 3 \rangle &=& \left [ \begin{array}{c}0\\0\\1\\0\end{array} \right ], &  \vert 4 \rangle &=& \left [ \begin{array}{c}0\\0\\0\\1\end{array} \right ].
  \end{array}
\end{equation}
In this basis the Hamiltonian is represented by
\begin{equation}
  H = \hbar \left [ \begin{array}{cccc}
      0 & \Omega^*_a/2 & 0 & 0 \\
      \Omega_a/2 & \delta_1 & \Omega_b/2 & 0 \\
      0 & \Omega^*_b/2 & \delta_2 & \Omega^*_c/2 \\
      0 & 0 & \Omega_c/2 & \delta_3
    \end{array} \right ].
  \label{moas:h}
\end{equation}
$\Omega_x$ are Rabi-frequencies that are defined as
\begin{equation}
  \begin{array}{rcl}
    \Omega_a &=& 2 g_a \sqrt{n_a}, \\ \Omega_b &=& 2 g_b \sqrt{n_b+1}, \\ \Omega_c &=& 2 g_c \sqrt{n_c}.
  \end{array} \label{moas:rabi}
\end{equation}
The form of the Hamiltonian (\ref{moas:h}) is particularly convenient since it is identical to that used in semi-classical calculations.  Writing the Hamiltonian in this way is possible since spontaneous emission has been neglected and each state evolves within a single resonant manifold.  Otherwise, transitions could be made to non-resonant manifolds and it would be insufficient to model each manifold in isolation.

\section{Non-degenerate perturbation theory in two variables}

We wish to derive the eigenstates of the four-level atom interacting with three electromagnetic fields in the N-configuration.  By doing so, we will show that the evolution of the perturbed ground state $\vert 1 \rangle$ gives rise to a cross-Kerr interaction.  The eigenstates of the Hamiltonian (\ref{moas:h}) will be calculated by using time-independent non-degenerate perturbation theory in two variables.

We begin by splitting the Hamiltonian into three parts: $\hat{H}_0, \hat{V}_a$ and $\hat{V}_c$.  $\hat{H}_0$ consists of the four energy levels, where the states $\vert 2 \rangle$ and $\vert 3 \rangle$ are coupled by the field $b$. This system represents an exactly solvable two-level subsystem with two additional uncoupled levels.  $\hat{V}_a$ is the weak coupling from $\vert 1 \rangle$ to $\vert 2 \rangle$ due to the field $a$ and $\hat{V}_c$ is the coupling between $\vert 3 \rangle$ and $\vert 4 \rangle$ produced by $c$.  The strength of the pertubations is parameterised by a single variable each: $\epsilon_a = \vert \Omega_a \vert / 2$ and $\epsilon_c = \vert \Omega_c \vert / 2$, whereast he structure is determined by the matrix operators $\hat{V}_a$ and $\hat{V}_c$.  This is depicted in the bare atomic basis by Fig. (\ref{int:fls}).  Therefore, the total Hamiltonian is,
\begin{equation}
  \hat{H}=\hat{H}_0 + \epsilon_a \hat{V}_a + \epsilon_c \hat{V}_c.
\end{equation}
The Hamilonain $\hat{H}_0$ of the two-level subsystem has the matrix representation
\begin{equation}
    H_0 = \left [ \begin{array}{cccc}0&0&0&0 \\ 0&\delta_1&\Omega_b/2&0 \\ 0&\Omega^*_b/2&\delta_2&0 \\ 0&0&0&\delta_3\end{array} \right ]. \label{npttv:h0}
\end{equation}
The two perturbations $\hat{V}_a$ and $\hat{V}_c$ have the representations
\begin{eqnarray}
    V_a &=& \left [ \begin{array}{cccc}0&e^{-i \phi_a}&0&0 \\ e^{i \phi_a}&0&0&0 \\ 0&0&0&0 \\ 0&0&0&0 \end{array} \right ], \label{npttv:va}\\
    V_c &=& \left [ \begin{array}{cccc}0&0&0&0 \\ 0&0&0&0 \\ 0&0&0&e^{-i \phi_c} \\ 0&0&e^{i \phi_c}&0 \end{array} \right ], \label{npttv:vc}
\end{eqnarray}
where $\phi_a = arg(\Omega_a)$ and $\phi_c = arg(\Omega_c)$.

To use perturbation theory we must first determine the eigenstates of the exactly solvable system $\hat{H}_0$.  These are given by the uncoupled states $\vert 1 \rangle$ and $\vert 4 \rangle$ and the dressed states of the two-level subsystem:
\begin{eqnarray}
  \vert \phi^{(0,0)}_1 \rangle &=& \vert 1 \rangle, \\
  \vert \phi^{(0,0)}_2 \rangle &=& \frac{1}{N_-} (\Omega_b \vert 2 \rangle + 2(\lambda_- - \delta_1) \vert 3 \rangle), \\  
  \vert \phi^{(0,0)}_3 \rangle &=& \frac{1}{N_+} (\Omega_b \vert 2 \rangle + 2(\lambda_+ - \delta_1) \vert 3 \rangle), \\
  \vert \phi^{(0,0)}_4 \rangle &=& \vert 4 \rangle,
\end{eqnarray}
The corresponding eigenenergies are given by
\begin{eqnarray}
  \lambda^{(0,0)}_1 &=& 0, \\
  \lambda^{(0,0)}_2 &=& \frac{1}{2} \left ( (\delta_1 + \delta_2) - \sqrt{(\delta_1 - \delta_2)^2 + \vert \Omega_b \vert^2} \right ),\\
  \lambda^{(0,0)}_3 &=& \frac{1}{2} \left ( (\delta_1 + \delta_2) + \sqrt{(\delta_1 - \delta_2)^2 + \vert \Omega_b \vert^2} \right ), \\
  \lambda^{(0,0)}_4 &=& 0.
\end{eqnarray}
Since the total Hamiltonian $\hat{H}$ differs from the unperturbed system only by the weak pertubations $\hat{V}_a$ and $\hat{V}_c$ we expect the eigenstates and eigenenergies of both systems to be similar.  To find approximate eigenstates and eigenergies of $\hat{H}$ we expand these in power series of the perturbation strengths $\epsilon_a$ and $\epsilon_c$.
\begin{eqnarray}
    E_n &=&\sum^{\infty}_{i, j=0} \epsilon^i_a \epsilon^j_c E^{(i,j)}_n, \label{npttv:eps}\\
    \vert \phi_n \rangle &=& \sum^{\infty}_{i,j=0} \epsilon^i_a \epsilon^j_c \vert \phi^{(i,j)}_n \rangle. \label{npttv:sps}
\end{eqnarray}
It is also convenient to expand the $\vert \phi^{(i,j)}_n \rangle$ coefficients in terms of the unperturbed Hamiltonian basis states
\begin{equation}
  \vert \phi^{(i,j)}_n \rangle = \sum^{4}_{s=1} a^{s(i,j)}_n \vert \phi^{(0,0)}_s \rangle. \label{npttv:uee}
\end{equation}
By substituting the power series (\ref{npttv:eps}) and (\ref{npttv:sps}) into the Sch\"{o}dinger Equation
\begin{equation}
  \left (  \hat{H}_0 + \epsilon_a \hat{V}_a + \epsilon_c \hat{V}_c \right ) \vert \phi_n \rangle = E_n \vert \phi_n \rangle,
\end{equation}
we find a set of equations that determine the eigenenergy corrections $E^{(i, j)}_n$ and eigenstate expansion coefficients $a^{s (i, j)}_n$. By using expansion (\ref{npttv:uee}) we find that the eigenenergy corrections are given by
\begin{equation}
  \mathop{\sum^{p,q}_{i,j=0}}_{(i,j) \neq (0,0)} E^{(i,j)}_n a^{n(p-i, q-j)}_n = \begin{array}[t]{l}\langle \phi^{(0,0)}_n \vert V_a \vert \phi^{(p-1, q)}_n \rangle + \\ \langle \phi^{(0,0)}_n \vert V_c \vert \phi^{(p, q-1)}_n \rangle. \end{array} \label{app:e}
\end{equation}
The eigenstate expansion coefficients are given by two different fomulae, depending on the indices $n$ and $m$ in the coefficient $a^{m (i, j)}_n$.  If $n \neq m$ then
\begin{widetext}
  \begin{equation}
    a^{m(p,q)}_n(E^{(0,0)}_n - E^{(0,0)}_m)=\langle \phi^{(0,0)}_m \vert V_a \vert \phi^{(p-1, q)}_n \rangle + \langle \phi^{(0,0)}_m \vert V_c \vert \phi^{(p, q-1)}_n \rangle - \mathop{\sum^{p,q}_{i=0, j=0}}_{(i,j) \neq (0,0)} E^{(i,j)}_n a^{m(p-i, q-j)}_n, \qquad (n \neq m). \label{app:a1}
  \end{equation}
\end{widetext}
From the normalisation of the wavefunction ($\langle \phi_n \vert \phi_n \rangle = 1$), we find a formula from which the $a^{n(i,j)}_n$ terms can be deduced:
\begin{equation}
  \sum^p_{i=0} \sum^q_{j=0} \sum^4_{s=1} a^{*s(i,j)}_n a^{s(p-i, q-j)}_n = 0. \label{app:a2}
\end{equation}
Using these formulae we can calculate approximate expressions for the four eigenstates of the total Hamiltonian.  However, only the perturbation caused to the ground state $\vert \phi^{(0,0)}_1 \rangle$ is of importance for our cross-Kerr scheme.  Under the influence of the strong coupling field ``b'' the system will relax into the radiatively stable ground state $\vert 1 \rangle$, since the other eigenstates all have populations in excited atomic levels.  Therefore, spontaneous emission will relax the system from any mixed state into this pure ground state.  We have assumed that the relaxation is caused by a weak coupling to the environment which acts only to drive the system into the ground state, but can then be neglected.  This is a common technique in non-equilibrium statistical mechanics.

To fourth-order, the eigenenergy of the state $\vert \phi_1 \rangle$ is given by
\begin{equation}
  \lambda_1 \approx \epsilon^2_a \lambda^{(2,0)}_1 + \epsilon^4_a \lambda^{(4,0)}_1 + \epsilon^2_a \epsilon^2_c \lambda^{(2,2)}_1,
\end{equation}
where the eigenvalue corrections are
\begin{eqnarray}
  \lambda^{(2,0)}_1 &=& - \frac{\delta_2 \vert \Omega_a \vert^2}{4 \delta_1 \delta_2 - \vert \Omega_b \vert^2}, \\
  \lambda^{(4,0)}_1 &=& \frac{\delta_2 (4 \delta^2_2 + \vert \Omega_b \vert^2) \vert \Omega_a \vert^4}{(4 \delta_1 \delta_2 - \vert \Omega_b \vert^2)^3}, \\ 
  \lambda^{(2,2)}_1 &=& -\frac{\vert \Omega_b \vert^2 \vert \Omega_a \vert^2 \vert \Omega_c \vert^2}{4 \delta_3 (4 \delta_1 \delta_2 - \vert \Omega_b \vert^2)^2}. \label{npttv:lambda}
\end{eqnarray}
Having calculated the eigenenergy of this state we easily find the stable state evolution to be described by
\begin{equation}
  \vert \phi_1 (t) \rangle = \exp(-i \lambda_1 t) \vert \phi_1(0) \rangle.
\end{equation}
Recalling the definition of the Rabi frequencies (\ref{moas:rabi}) we may express each term of (\ref{npttv:lambda}) in terms of photon numbers.  Moreover, since the eigenstate $\vert \phi_1 \rangle$ is given to zeroth order by $\vert 1, n_a, n_b, n_c \rangle$ then we can replace the photon numbers with their corresponding photon-number operators, when acting on this state.  This is possible because the state $\vert 1, n_a, n_b, n_c \rangle$ is an eigenstate of the photon-number operators $\hat{n}_x$ with eigenvalues $n_x$.  Therefore
\begin{equation}
  \vert \phi_1 (t) \rangle = \exp \left (-i \left \{L \hat{n}_a + S \hat{n}^2_a + K \hat{n}_a \hat{n}_c \right \} t \right ) \vert \phi_1(0) \rangle,
\end{equation}
where we have defined
\begin{eqnarray}
  L &=& - \frac{\delta_2 \vert g_a \vert^2}{\delta_1 \delta_2 - \vert g_b \vert^2(n_b + 1)},\\
  S &=& \frac{\delta_2 ( \delta_2^2 + \vert g_b \vert^2 (n_b+1)) \vert g_b \vert^4}{(\delta_1 \delta_2 - \vert g_b \vert^2 (n_b+1))^3},\\
  K &=& \frac{- \vert g_a \vert^2 \vert g_b \vert^2 \vert g_c \vert^2 (n_b + 1)}{\delta_3(\delta_1 \delta_2 - \vert g_b \vert^2 (n_b+1))^2}.
\end{eqnarray}
The coupling strengths $g_x$ and multi-photon detunings are defined in equations (\ref{moas:deltas}) and (\ref{moas:g}) respectively.

The evolution of the ground state $\vert 1, n_a, n_b, n_c \rangle$ is clearly generated by an effective Hamiltonian of the form
\begin{equation}
  \hat{H}_{eff} = \hbar \left( L \hat{n}_a + S \hat{n}^2_a + K \hat{n}_a \hat{n}_c \right ).
  \label{npttv:heff}
\end{equation}
The coefficients $L, S$ and $K$ represent the linear, self-Kerr and cross-Kerr responses of the atom.  The linear and self-Kerr energy contributions are due to the the fields ``a'' and ``b'' coupling between the states $\vert 1 \rangle, \vert 2 \rangle$ and $\vert 3 \rangle$: this consistutes a $\lambda$ subsystem.  However, it is known that when the two-photon transition from $\vert 1 \rangle$ to $\vert 3 \rangle$ is resonant ($\delta_2=0$), the linear and self-Kerr terms vanish.  This is due to the generation of a {\it dark} eigenstate of the $\lambda$ subsystem.  The {\it darkstate} consists of a superposition of the lower atomic levels $\vert 1 \rangle$ and $\vert 3 \rangle$, with no population in the excited state $\vert 2 \rangle$.  Since there are no material polarisations between the $\vert 1 \rangle \leftrightarrow \vert 2 \rangle$ or $\vert 2 \rangle \leftrightarrow \vert 3 \rangle$ transitions the state is non-interacting, or {\it dark}, to the applied fields.  Therefore the linear and self-Kerr responses vanish ($L=S=0$).

Furthermore, when $\delta_2=0$ then the field ``c'' coupling level $\vert 3 \rangle$ of the $\lambda$ subsystem to $\vert 4 \rangle$ perturbs the darkstate.  The perturbation is due to an adiabatic Stark shift of the atomic level $\vert 3 \rangle$ by the field ``c''.  Within the two-photon resonance regime, we find that the evolution of the system reduces to a pure cross-Kerr evolution.  The terms $L, S, $ and $K$ are given by
\begin{eqnarray}
  L &=& 0, \\
  S &=& 0, \\
  K &=& - \frac{\vert g_a \vert^2 \vert g_c \vert^2}{\delta_3 \vert g_b \vert^2 ( \hat{n}_b + 1)}.
\end{eqnarray}
The effective Hamiltonian ({\ref{npttv:heff}) then reduces to the form of a pure cross-Kerr interaction (\ref{int:cke}) between the fields ``a'' and ``c'':
\begin{equation}
  \hat{H}_{eff} = \hbar K \hat{n}_a \hat{n}_c.
\end{equation}

\section{Electrical Susceptibility}
We now calculate the absorption that accompanies the linear, Kerr and cross-Kerr responses of the atom.  To do so, we consider the macroscopic material polarisation at the frequency of both the probe fields.
\begin{equation}
  P(t) = \frac{1}{2} \sum_{n=\{a,c\}} P(\omega_n) e^{-i \omega_n t} + P^*(\omega_n) e^{i \omega_n t}.
  \label{es:p}
\end{equation}
For instance, we expect that the component of polarisation at the frequency $\omega_a$ will display linear, self-Kerr and cross-Kerr contributions.  This is described by
\begin{widetext}
  \begin{equation}
    P(\omega_a) \approx \epsilon_0 \chi^{(1)}(\omega_a; \omega_a) E(\omega_a) + \frac{3}{4} \epsilon_0 \chi^{(3)}(\omega_a; \omega_a, -\omega_a, \omega_a) \vert E(\omega_a) \vert^2 E(\omega_a) + \frac{3}{2} \epsilon_0 \chi^{(3)}(\omega_a; \omega_c, -\omega_c, \omega_a) \vert E(\omega_c) \vert^2 E(\omega_a) + \dots
  \end{equation}
\end{widetext}
We note that the electric fields are related to the Rabi-frequencies by,
\begin{equation}
  E(\omega_a) = -\frac{\Omega_a \epsilon_a}{g_a}, \qquad E(\omega_c) = - \frac{\Omega_c \epsilon_c}{g_c}.
\end{equation}
Now, we expect that the macroscopic polarisation described by equation ({\ref{es:p}) will be related in some way to the microscopic state of the atom.  We begin by constructing the wavefunction of the $\vert \phi_1 \rangle$ eigenstate up to third-order by using the expansion (\ref{npttv:sps}).  The density matrix of this pure eigenstate is
\begin{equation}
  \rho = \vert \phi_1 \rangle \langle \phi_1 \vert.
\end{equation}
In particular, the atom interacts with the electromagnetic fields via the off-diagonal elements of the density matrix.  In the interaction picture these are time-independent and are given by
\begin{eqnarray}
  \tilde{\rho}_{21} &=& \langle 2 \vert \rho \vert 1 \rangle, \\
  \tilde{\rho}_{43} &=& \langle 4 \vert \rho \vert 3 \rangle.
\end{eqnarray}
So far the evolution of the system has been described by the Schr\"{o}dinger equation, and is therefore unitary.  To model the atomic absorption we require that the detunings of the electromagnetic fields from the atomic resonances become complex \cite{NLO}.
\begin{eqnarray}
  \delta_1 &\rightarrow& \delta_1 - i \gamma_1, \\
  \delta_2 &\rightarrow& \delta_2 - i \gamma_2, \\
  \delta_3 &\rightarrow& \delta_3 - i \gamma_3.
\end{eqnarray}
It is now possible to equate the material polarisation with that described by the off-diagonal density matrix elements in the Sch\"{o}dinger picture.  We say that
\begin{equation}
  P(t) = \rho_{12} p_{21} + \rho_{34} p_{43} + c.c.,
\end{equation}
where $p_{ij}= e \langle i \vert {\b r} \vert j \rangle$ are the dipole matrix elements.  By Taylor expanding the off-diagonal density matrix elements in terms of the fields $\Omega_a$ and $\Omega_c$ we may find expressions for the linear,  self-Kerr and cross-Kerr electrical susceptibilities.  For the $\vert 1 \rangle \leftrightarrow \vert 2 \rangle$ transition we find that these are given by
\begin{widetext}
  \begin{equation}
    \chi^{(1)}(\omega_a; \omega_a) = \frac{\hbar \vert g_a \vert^2 (i \gamma_2 - \delta_2)}{\epsilon_0 \epsilon^2_a (\gamma_1 + i \delta_1)(\gamma_2 + i \delta_2) + \vert g_b \vert^2 (n_b+1)]},
  \end{equation}
  \begin{equation}
    \chi^{(3)}(\omega_a; \omega_a, -\omega_a, \omega_a) = \frac{2 \hbar \vert g_a \vert^4 (i \gamma_2 - \delta_2)[(\gamma_2 + i \delta_2)^2 - \vert g_b \vert^2 (n_b + 1)]}{3 \epsilon_0 \epsilon^4_a [(\gamma_1 + i \delta_1)(\gamma_2 + i \delta_2) + \vert g_b \vert^2 (n_b + 1)]^3},
  \end{equation}
  \begin{equation}
    \chi^{(3)}(\omega_a; \omega_c, -\omega_c, \omega_a) = \frac{\hbar \vert g_a \vert^2 \vert g_b \vert^2 \vert g_c \vert^2 (n_b +1)}{6 \epsilon_0 \epsilon^2_a \epsilon^2_c (-i \gamma_3 + \delta_3)[(\gamma_1 + i \delta_1)(\gamma_2 + i \delta_2) + \vert g_b \vert^2 (n_b+1) ]^2},
  \end{equation}
\end{widetext}
For the $\vert 3 \rangle \leftrightarrow \vert 4 \rangle$ transition we expect only a cross-phase modulation to occur.  Indeed we find that
\begin{equation}
  \chi^{(3)}(\omega_c; \omega_a, - \omega_a, \omega_c) = \chi^{(3)}(\omega_a; \omega_c, -\omega_c, \omega_a).
\end{equation}
The real and imaginary parts of the susceptibilities for the $\vert 1 \rangle \leftrightarrow \vert 2 \rangle$ and $\vert 3 \rangle \leftrightarrow \vert 4 \rangle$ transitions are plotted in figures (\ref{es:x1}), (\ref{es:x3s}) and (\ref{es:x3c}).  Here we split each susceptibility into real and imaginary parts
\begin{equation}
  \chi^{(n)} = \chi^{'(n)} + i \chi^{''(n)},
\end{equation}
where $\chi^{'(n)}$ and $\chi^{''(n)}$ are real.  The linear, self-Kerr and cross-Kerr susceptibility are denoted by $\chi^{(1)}, \chi^{(3)}_s$ and $\chi^{(3)}_c$.

\begin{figure}[h]
  \center
  \includegraphics[width=7.5cm]{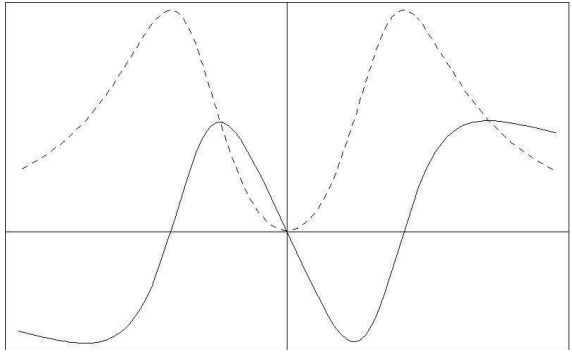}
  \caption{The real $\chi^{'(1)}$ (solid line) and imaginary $\chi^{''(1)}$ (dashed line) components of the linear susceptibility experienced by the field ``a'', plotted versus the detuning $\Delta_a$.}
  \label{es:x1}
\end{figure}
Figures (\ref{es:x1}) and (\ref{es:x3s}) show the linear and self-Kerr susceptibilities experienced by the field ``a'' for various detunings, given a resonant control field ``b''.  These graphs illustrate a normal electromagnetically induced transparency response of the lambda subsystem.  For exact resonance of the probe field $\Omega_a$ the lambda subsystem forms an atomic {\it dark state} and the real and imaginary parts of the susceptibilities vanish.
\begin{figure}[b]
  \center
  \includegraphics[width=7.5cm]{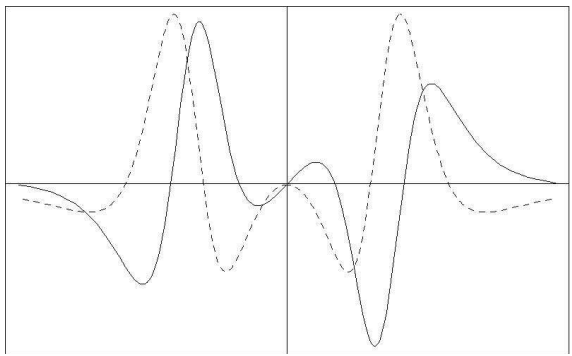}
  \caption{The real $\chi^{'(3)}_s$ (solid line) and imaginary $\chi^{''(3)}_s$ (dashed line) components of the self-Kerr susceptibility experienced by the field ``a'', plotted versus the detuning of $\Delta_a$.}
  \label{es:x3s}
\end{figure}

The final graph (\ref{es:x3c}) shows the dependence of the cross-Kerr susceptibility on the detuning of the probe field ``c'' for exact resonance of the fields ``a'' and ``b''.  When the field ``c'' is resonant, then the cross-Kerr interaction vanishes since the real part of $\chi^{(3)}_c$ is zero.  However, an absorption peak remains.  Off-resonance we find that the real susceptibility initially increases and then decays slowly.  Moreover, the real susceptibility decays much more slowly than the imaginary (absorptive) part.  Therefore, when the field ``c'' is off-resonance, then we achieve a non-zero cross-Kerr interaction with a rapidly vanishing absoption.
\begin{figure}[h]
  \includegraphics[width=7.5cm]{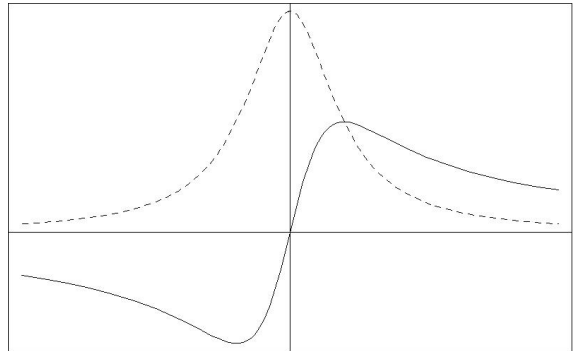}
  \caption{The real $\chi^{'(3)}_c$ (solid line) and imaginary $\chi^{''(3)}_c$ (dashed line) components of the cross-Kerr susceptibility experienced by the fields ``a'' and ``c'', plotted versus the detuning $\Delta_c$.}
  \label{es:x3c}
\end{figure}

\section{Conclusions}
We have shown that the four-level atom in the N-configuration can give rise to a cross-Kerr type effective Hamiltonian.  However, as a development on earlier work \cite{CKIFLAS} there are no restrictions on the detunings of the electromagnetic fields.  Importantly, this enables us to show how the strength of the cross-Kerr interaction depends on the detunings of all three fields.  Moreover, we also show that in general the cross-Kerr interaction will be accompanied by a linear and self-Kerr response of the atom.  Only when the fields ``a'' and ``b'' forming the $\lambda$ subsystem are Raman-resonant do we achieve a pure cross-Kerr interaction.

In addition to fully exploring the unitary aspects of the evolution, we also derive the absorption that will accompany these interactions.  We achieve this by calculating the complex electrical susceptibilities for the weak-probe transitions.  It is shown that if the two fields forming the $\lambda$ subsystem are Raman-resonant then the absorptive component of the linear and self-Kerr susceptibilities will also vanish.  However, the absorption related to the cross-Kerr interaction will remain.  Nonetheless, by increasing the detuning of the weak probe field $\Omega_c$ one can achieve a slowly decreasing cross-Kerr interaction whilst the absorptive contribution of the cross-Kerr susceptibility rapidly vanishes.  Remarkably, in this regime one can produce a large cross-Kerr interaction with vanishing absorption in the four-level atom.

\section{Acknowledgments}
This work was supported by EPSRC, SUPA, and the European Union COVAQIAL project.

\bibliography{references}
\bibliographystyle{unsrt}

\end{document}